\documentclass[preprint,epsfig,onecolumn,floats,showpacs]{revtex4}
\usepackage{amssymb}
\usepackage{graphicx}
\usepackage{epsfig}
\usepackage{amsmath}
\usepackage{amsfonts}

\begin{document}

\title{Dispersive Coupling Between the Superconducting Transmission Line
Resonator and the Double Quantum Dots}
\author{Guo-Ping Guo}
\email{gpguo@ustc.edu.cn}
\author{Hui Zhang}
\author{Yong Hu}
\author{Tao Tu}
\email{tutao@ustc.edu.cn}
\author{Guang-Can Guo}
\affiliation{Key Laboratory of Quantum Information, Chinese Academy
of Sciences, University of Science and Technology of China, Hefei,
Anhui 230026, China}

\begin{abstract}
Realization of controllable interaction between distant qubits is
one of the major problems in scalable solid state quantum computing.
We study a superconducting transmission line resonator (TLR) as a
tunable dispersive coupler for the double-dot molecules. A general
interaction Hamiltonian of  $n$ two-electron spin-based qubits and
the TLR is presented, where the double-dot qubits are biased at the
large detuning region and the TLR is always empty and virtually
excited. Our analysis o the main decoherence sources indicates that
various major quantum operations can be reliably implemented with
current technology.
\end{abstract}

\maketitle


Semiconductor quantum dot is one of the most promising candidates
for quantum computation \cite{Loss,Hanson}. With the help of
external electromagnetic bias, one can coherently manipulate the
quantum state of the trapped electron \cite{Engel,Petta,Taylor1}.
Recently, the singlet and triplet states of the two electrons in a
double-dot molecule have been proposed as logical qubit
\cite{Taylor2}. Compared with the single electron qubit, this
two-electron spin-based qubit has proved to be more robust against
noise from the quasi-static nuclear spins \cite{Petta} as well as
low frequency fluctuation of the electrostatic gates \cite{Taylor3}.

Stimulated by these advances, we focus on the further steps towards
realizing scalable quantum computing in this solid state system. One
of the critical next steps is to implement two-qubit quantum gates.
Capacitive coupling, i. e. directly coupling via Coulomb interaction
of electrons in quantum dots is the most straightforward way
\cite{Burkard1,Burkard2,Guo1,Guo2}. However, it fails for distant
qubits because the Coulomb interaction decays rapidly as the
distance between the qubits increases. Moreover, even if one could
place many qubits in a small area to obtain enough coupling
strength, problems such as the residual coupling arise, hence the
design of quantum gates would become more and more complicated. All
these issues make alternative coupling methods desirable for the
scalable quantum computing in the future.

In the very latest papers, the idea of combining quantum dots with
the circuit QED system
\cite{CircuitQEDPRA,CircuitQEDNature,CircuitQEDPRL} is attracting
more and more attention. Ref. \cite{Taylor3} and Ref.
\cite{Childress} have suggested that the charge degree of freedom of
the double-dot and the quantized voltage of a transmission line
resonator (TLR) can be coupled via a capacitor. These works
suggested that in the resonant strong coupling limit, the energy and
the quantum information can be transferred between the double-dot
and the TLR. However, only one double-dot qubit is considered to
interact with the TLR in these previous works Ref.
\cite{Taylor3,Childress}. Therefore, it is nontrivial to find a
general interaction formalism involving $n$ double-dot qubits and
the TLR, which can be exploited to realized various important
quantum information processes (QIP) such as control-not gate between
any double-dot qubits and multi-qubit entanglement generation.

In this letter, instead of the resonant coupling of one double-dot
qubit with TLR in previous papers, we consider the dispersive
operations \cite{SBZheng} of multi double-dot molecules and circuit
QED combined system. A general interaction Hamiltonian of $n$
double-dot qubits and the TLR is presented. These two-electron
spin-based qubits are biased at the large detuning region and the
qubit-TLR interaction can be tuned on and off by the control of the
qubit level splitting. Since the TLR is always empty and only
virtually excited, the quantum coherence of this system is immune to
the energy loss of the cavity. This scalable architecture is
flexible enough to allow for various major quantum operations such
as nontrivial multi-qubit quantum gates and entanglement generation.
Since long life time for the double-dot qubit has already been
achieved, the proposed quantum gates could be realized with very
high fidelity.

Let us illustrate our idea intuitively. The system we study is a TLR
of length $L$, with inductance $F$ and capacitance $C$ per unit
length, coupled to $n$ double-dot qubits by coupling capacitors
$C_{ci}$ and to external input/output by wiring capacitor $C_{0}$
from left and right, as shown in Fig. (1). In the following, we
treat the TLR and the individual qubits as independent systems with
the small coupling as perturbation \cite{HuangRSThesis}.

\begin{figure}[tbh]
\epsfig{file=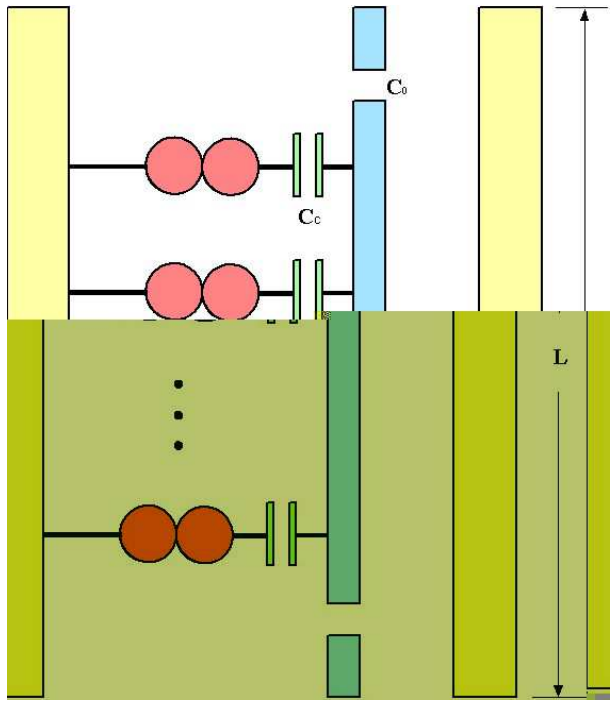,width=8cm} \caption{(Color Online) Coupled
system of a TLR and $n$ double-dot molecules. The length of the TLR
is L (about ten millimeters). $C_{0}$ is the wiring capacitor
between TLR and external input/output. The double-dot qubits are
connected to the TLR via the coupling capacitors $C_{ci}$. The TLR
acts as a data bus between the double-dot qubits.}
\end{figure}

\emph{The TLR}\textit{\ }-----A single TLR can be well described by
a series of inductors with each node capacitively connected to the
ground, as shown in Fig. (2). To transmit input and output signals,
the TLR should be coupled to the external coil by a wiring capacitor
$C_{0}$. Denoting $\epsilon _{0}=C_{0}/LC$ and focusing only on the
full wave mode, we can get the Hamiltonian of the TLR
\begin{equation}
H_{\mathrm{cavity}}=\hbar \omega a^{\dagger }a,
\end{equation}
where $a$ is the annihilation operator of the full wave mode. The
frequency of this mode is slightly renormalized by the wiring
capacitor as $\omega
\thickapprox \omega _{0}[1-2\epsilon _{0}]$ where $\omega _{0}=2\pi /\left( L%
\sqrt{FC}\right) $. The voltage distribution inside the TLR has the
form
\begin{equation}
V_{TLR}(x)=\sqrt{\hbar \omega /LC}(a^{\dagger }+a)\cos [kx+\delta ],
\label{VTLRPHYS}
\end{equation}
where the small phase shift $\delta $ satisfies the condition $\tan
\delta =2\pi \epsilon _{0}$.

\begin{figure}[tbh]
\epsfig{file=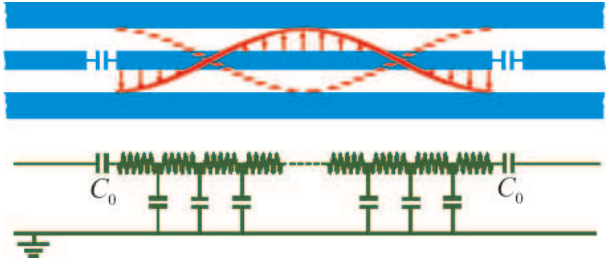,width=8cm} \caption{(Color Online) Circuit
representation of a TLR. The TLR consists of a full-wave section of
superconducting coplanar waveguide. The red line represents the full
wave mode. The coupling capacitors connected to the input/output
wiring slightly modify the frequency and phase of the TLR's modes.}
\end{figure}

\emph{The double-dot qubits }-----As shown in Fig. (3), a double-dot
molecule is formed by a layer of two dimensional electron gas
restrained by several electrostatic gates used to control the
potentials of individual dots and the interdot tunneling. Its low
energy spectrum is plotted in Fig. (4). Due to the Coulomb
interaction and Pauli principle, near the \textquotedblleft
saddle\textquotedblright\ point (gray area in Fig. (4)) the molecule
can be reduced to an artificial three-level system with the
Hamiltonian

\begin{align}
H_{DD}& =E_{T}\left\vert (1,1)T_{0}\right\rangle \left\langle
(1,1)T_{0}\right\vert +E_{S}\left\vert \left( 1,1\right) S\right\rangle
\left\langle \left( 1,1\right) S\right\vert -\varepsilon \left\vert \left(
0,2\right) S\right\rangle \left\langle \left( 0,2\right) S\right\vert
\label{PhysicHamilDD} \\
& +T_{C}\left( \left\vert \left( 1,1\right) S\right\rangle \left\langle
\left( 0,2\right) S\right\vert +\left\vert \left( 0,2\right) S\right\rangle
\left\langle \left( 1,1\right) S\right\vert \right) ,  \notag
\end{align}%
where the notation $\left( n_{l},n_{r}\right) $ indicates $n_{l}$
electrons on the "left"\ dot and $n_{r}$ electrons on the "right"\
dot, $S$ and $T$ represent spin singlet and triplet states,
$\varepsilon $ and $T_{C}$ denote the external voltage bias and
tunneling amplitude between two dots. It is a good approximation to
set $E_{T}\approx E_{S}=0$ \cite{Taylor1}, therefore Eq.
(\ref{PhysicHamilDD}) has eigenstates
\begin{eqnarray}
\left\vert \widetilde{S}\right\rangle &=&\cos \theta \left\vert \left(
1,1\right) S\right\rangle +\sin \theta \left\vert (0,2)S\right\rangle ,
\label{EigenStateDD} \\
\left\vert \widetilde{G}\right\rangle &=&-\sin \theta \left\vert
\left( 1,1\right) S\right\rangle +\cos \theta \left\vert
(0,2)S\right\rangle ,\notag
\end{eqnarray}%
here the parameter $\theta $ rests on the external bias $\varepsilon
$ and $T_{C}$ \cite{Taylor1}. The voltage
of the left dot is%
\begin{equation}
V_{dot}=\frac{2e}{C_{tot}}\left\vert (0,2)S\right\rangle \left\langle
(0,2)S\right\vert +\frac{e}{C_{tot}}\left\vert (1,1)S\right\rangle
\left\langle (1,1)S\right\vert ,  \label{VDOTPHY}
\end{equation}%
where $C_{tot}$ is the total capacitance of the double-dot molecule.
In realistic experiments, the choice $\theta =\pi /4$ is favorable
for quantum operation because at this point the energy
difference between $\left\vert \widetilde{S}%
\right\rangle $ and \ $\left\vert \widetilde{G}\right\rangle $ is
insensitive to the first-order fluctuation of $\varepsilon $
\cite{Vion} and the nuclear hyperfine field \cite{Taylor3}. Therefore, we denote $%
\left\vert 0\left( 1\right) \right\rangle =\left( \left\vert \downarrow
\uparrow \right\rangle \mp \left\vert (0,2)S\right\rangle \right) /\sqrt{2}$
and rewrite Eq. (\ref{VDOTPHY}) as%
\begin{equation}
V_{dot}=e\left( I+\sigma _{x}\right) /2C_{tot}  \label{VDOTPAULI}
\end{equation}%
in the basis $\left\{ \left\vert 0\right\rangle ,\left\vert 1\right\rangle
\right\} $.

\begin{figure}[tbh]
\epsfig{file=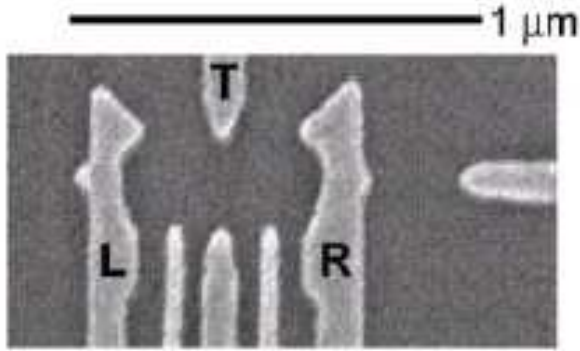,width=8cm} \caption{Micrograph of a double-dot
sample. This figure is reporduced from Fig.1 of Ref.[4]. The L and R
gate represent the left and right gate respectively in the text, and
the T gate can be used to adjust the tunneling coupling strength
between two dots.}
\end{figure}

\begin{figure}[tbh]
\epsfig{file=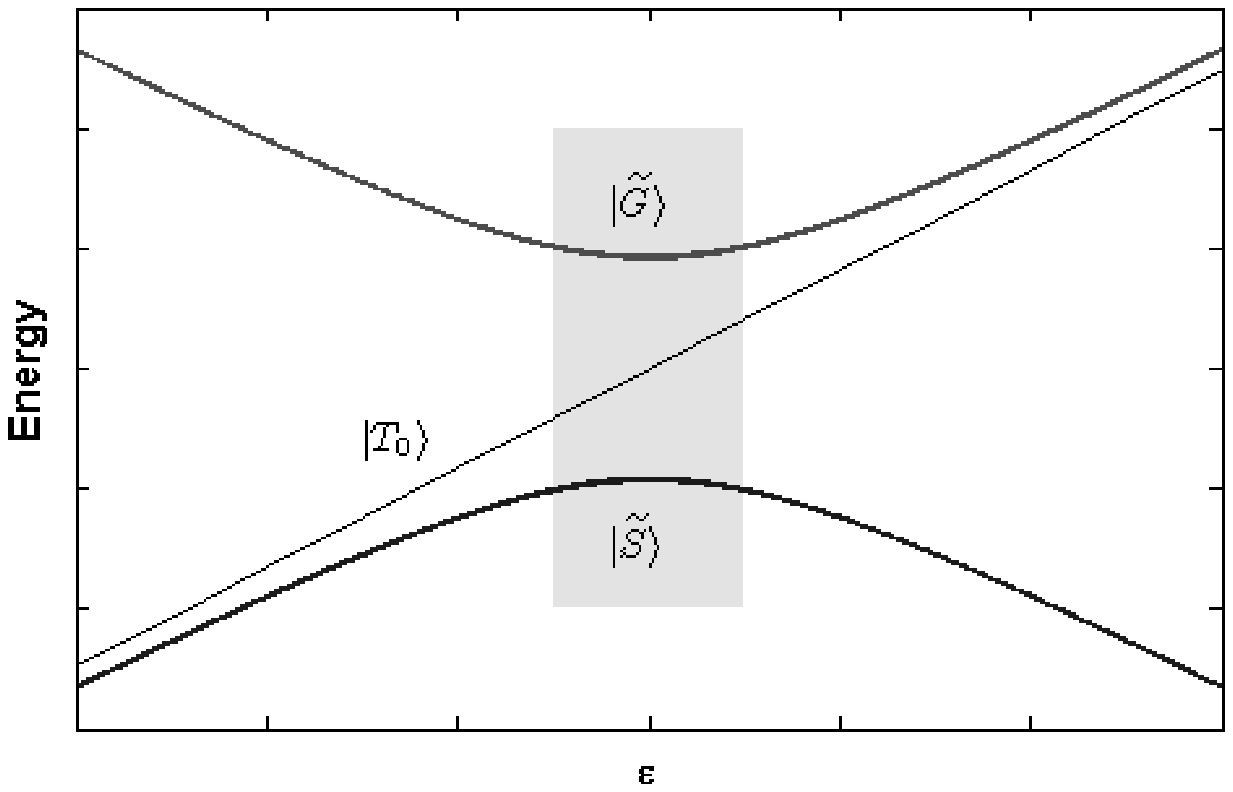,width=8cm} \caption{Lowest-energy states of
the double-dot molecule. The gray regions is the "saddle" point in
the text.}
\end{figure}

\emph{The combined system}\textit{\ }-----As shown in Fig. (1), when
a double-dot is coupled to a TLR with a capacitor $C_{c}$, the
charging energy of $C_{c}$ provides a qubit-resonator interaction
\begin{equation}
H_{int}=C_{c}V_{TLR}(x)V_{dot}.
\end{equation}%
With Eq. (\ref{VTLRPHYS}) and Eq. (\ref{VDOTPAULI}), we can get the
standard Jaynes-Cumming (JC) model
\begin{equation}
H_{int}=g(x)\left( a\sigma ^{+}+a^{+}\sigma ^{-}\right) ,
\end{equation}%
where $\sigma ^{+}=\left\vert 1\right\rangle \left\langle 0\right\vert $, $%
\sigma ^{-}=\left\vert 0\right\rangle \left\langle 1\right\vert $, and the
coupling $g$ factor
\begin{equation}
g\left( x\right) =\frac{eC_{c}}{C_{tot}}\sqrt{\frac{\hbar \omega
}{LC}}\cos [kx+\delta ].
\end{equation}%
It is interesting to note the single dot qubit which exploits the
spin degree of freedom of the trapped electron can not be coupled to
the TLR because of its charge degeneracy.

Previous papers have suggested to put the double-dot qubits on the
two leads of the cavity and to operate the circuit QED system
resonantly. Alternatively, we consider putting $n$ double-dot
molecules inside the cavity and operating them in the region of
large detuning. As shown in Fig. (1), there are $n$ double-dot
molecules coupled to a TLR. Since the TLR length is in the range of
$10$ $\mathrm{mm}$, we can place the qubit far enough from each
other, thus the Coulomb interaction between distant qubits can be
neglected. In the interaction picture, we can get an effective
Hamiltonian of $n$ double-dot molecules and the TLR as

\begin{equation}
H_{int}=\sum_{j=1}^{n}g_{j}\left( e^{-i\tau _{j}t}a^{+}\sigma
_{j}^{-}+e^{i\tau _{j}t}a\sigma _{j}^{+}\right) ,\text{
}j=1,2,\centerdot \centerdot \centerdot ,n.  \label{HamiPhYs}
\end{equation}%
where $\tau _{j}$ is the detuning between the transition frequency
of qubit  $j$ and the full mode frequency of the TLR. When the
condition of dispersive detuning, i. e. $\tau _{j}\gg g_{j}$ for
$j=1,2,\centerdot \centerdot \centerdot ,n$, is fulfilled, the
Hamiltonian Eq. (\ref{HamiPhYs}) can be transformed to \cite{guo3}

\begin{equation}
H_{eff}=\lambda\left[ \sum_{i,j=1}^{n}\left(
\sigma_{j}^{+}\sigma_{i}^{-}aa^{+}-\sigma_{j}^{-}\sigma_{i}^{+}a^{+}a\right) %
\right] ,  \label{HamiEffe}
\end{equation}
where $\lambda=g^{2}/\tau$ (for simplicity we assume identical
coupling strength $g$ and detuning $\tau$).

Various important quantum information processes could be realized
with the above Hamiltonian Eq. (\ref{HamiEffe}). As an example, we
show how to perform a CPHASE gate between any two double-dot qubits.
Initially we prepare the cavity field in the vacuum state and the
two target qubits in the state $\left\vert 1\right\rangle
_{1}\left\vert 0\right\rangle _{2}$ \cite{SBZheng}.  As the
effective coupling strength $\lambda $ depends on the detuning $\tau
$, we can selectively tune the two target qubits into the dispersive
region by choosing a relatively small (but still dispersive)
detuning while retain other qubits decoupled. In this circumstance,
the reduced form of $H_{eff}$ in the basis $\left\{ \left\vert
00\right\rangle ,\left\vert 10\right\rangle ,\left\vert
01\right\rangle ,\left\vert 11\right\rangle \right\} $ is

\begin{equation}
H_{20}=\lambda\left[ \sum_{j=1,2}^{n}\left\vert 1\right\rangle
_{jj}\left\langle 1\right\vert +\left(
\sigma_{1}^{+}\sigma_{2}^{-}+\sigma_{1}^{-}\sigma_{2}^{+}\right) \right] ,
\end{equation}
and the evolution operator is

\begin{align}
U\left( t\right) & =\exp(-itH_{20})  \notag \\
& =\left(
\begin{array}{cccc}
1 & 0 & 0 & 0 \\
0 & \frac{e^{-2i\lambda t}+1}{2} & \frac{e^{-2i\lambda t}-1}{2} & 0 \\
0 & \frac{e^{-2i\lambda t}-1}{2} & \frac{e^{-2i\lambda t}+1}{2} & 0 \\
0 & 0 & 0 & \frac{e^{-2i\lambda t}}{2}%
\end{array}
\right) .
\end{align}
After a period $t_{0}=\pi/4\lambda$, the initial state $\left\vert
1\right\rangle _{1}\left\vert 0\right\rangle _{2}$ evolves into the
maximally entangled Einstein-Podolsky-Rosen (EPR) state

\begin{equation}
\left\vert \Psi_{EPR}\right\rangle =\frac{1}{\sqrt{2}}\left( \left\vert
1\right\rangle _{1}\left\vert 0\right\rangle _{2}-i\left\vert 0\right\rangle
_{1}\left\vert 1\right\rangle _{2}\right) ,
\end{equation}
where we have omitted the common phase factor $\pi/4$.

Now we discuss the decoherence problem resulting from the
environmental and systematical error, which is the main obstacle of
implementing quantum computing in solid state system. The
dissipation of the single TLR occurs mainly through coupling to the
external leads. In general the magnitude of this process can be
described by the decay factor $\kappa =\omega /Q$, where $Q$ is the
quality factor of the TLRs \cite{HuangRSThesis}. In the reported
high-finesse TLR cavity with eigenfrequency $\omega _{0}/2\pi =10$ $\mathrm{%
GHz}$ and $Q=1\times 10^{5}$, the $\kappa $ factor is of the order $0.1$ $%
\mathrm{MHz}$ \cite{CircuitQEDNature}. Moreover, in the dispersive
coupling condition, the TLR is always empty and only virtually
excited. The influence of the cavity loss is neglectable even when
the quality factor of TLR is lower, which greatly simplifies the
experimental implementation.

For the double-dot qubit, things become more complicated. Phonon in
the substrate as well as the finite impedance of the voltage bias
can cause fluctuations in the voltage bias and induce relaxation to
the qubits. This relaxation process can be suppressed by electrical
circuit engineering \cite{Nazarov1,Nazarov2} and very long
spin-relaxation time $T_{1}$ have already been observed
experimentally \cite{Kouwenhoven}. Another main noise source is the
nuclear spins. The fluctuations of the nuclear spins in the
substrate can lead to phase randomization of the electron spin via
hyperfine interaction. Nevertheless, the speed of nuclear field
variation is much slower than that of the electron spin evolution.
Due to the special singlet and triplet states encoding strategy,
this kind of decoherence can be also greatly reduced \cite{Taylor1}.
By further exploring spin
echo technique, dephasing time $%
T_{2}$ $\approx 1$ $\mathrm{\mu s}$ has been shown experimentally.

We should also mention the low frequency fluctuations of the
electrostatic bias. This noise, typically with a $1/f$ spectrum in
its low frequency part, is believed to be mainly produced by the
charge defect of the gate electrodes. The gate bias of the qubit
drifts randomly when a electron tunnel in or out of the metallic
electrode. Due to the low frequency property, the effect of the
$1/f$ noise on the qubit is dephasing rather than relaxation. In
the recent experiments, the optimal point control is widely used to fight against the $%
1/f$ noise \cite{Vion}. By operating the qubit at the saddle point
where the linear longitudinal qubit-noise coupling vanishes (i. e. $%
\theta =\pi /4$ of our proposal), one can prolong the dephasing time by
several orders \cite{Martinis}.

Following the standard quantum theory of damping, we investigate the
combined influence of all the above decoherence processes on the coupled
system. After tracing out the reservoir degrees of freedom, we obtain the
master equation for the reduced density matrix $\rho $ of the three-party
system%
\begin{align*}
{\frac{d\rho }{dt}}& =-i[H_{\mathrm{int}}^{\mathrm{eff}},\rho ]+{\frac{%
\gamma _{\varphi 1}}{2}}\left[ \sigma _{z1}\rho \sigma _{z1}-\rho \right] \\
& +\left( {\frac{\gamma _{1}}{4}}\right) \left[ \sigma _{1}^{-}\rho \sigma
_{1}^{\dagger }-{\frac{1}{2}}\sigma _{1}^{\dagger }\sigma _{1}^{-}\rho -{%
\frac{1}{2}}\rho \sigma _{1}^{\dagger }\sigma _{1}^{-}\right] \\
& +{\frac{\gamma _{\varphi 2}}{2}}\left[ \sigma _{z2}\rho \sigma _{z2}-\rho %
\right] \\
& +\left( {\frac{\gamma _{2}}{4}}\right) \left[ \sigma _{2}^{-}\rho \sigma
_{2}^{\dagger }-{\frac{1}{2}}\sigma _{2}^{\dagger }\sigma _{2}^{-}\rho -{%
\frac{1}{2}}\rho \sigma _{2}^{\dagger }\sigma _{2}^{-}\right] ,
\end{align*}%
where $\gamma _{\varphi i}$ and $\gamma _{i}$\ are the pure
dephasing rate and relaxation rate of individual qubits. Choosing
$\gamma _{\varphi }=\gamma _{\varphi 1}=\gamma _{\varphi 2}$ and
$\gamma =\gamma _{1}=\gamma
_{2}$, we calculate the error probability $D$\ of the EPR generation versus $%
\gamma _{\varphi }$ and $\gamma $, as shown in Fig. 4. In this
calculation we choose $g/2\pi =100$ $\mathrm{MHz}$ and the large
detuning $\tau =10g$ \cite{Taylor3}. This simulation shows that the
error probability $D$ can be lower than $1\%$ when we used the
previous reported decoherence rates $\gamma /2\pi =0.2$
$\mathrm{MHz}$ and $\gamma _{\varphi }/2\pi =0.5$ $\mathrm{MHz}$,
\begin{figure}[tbh]
\epsfig{file=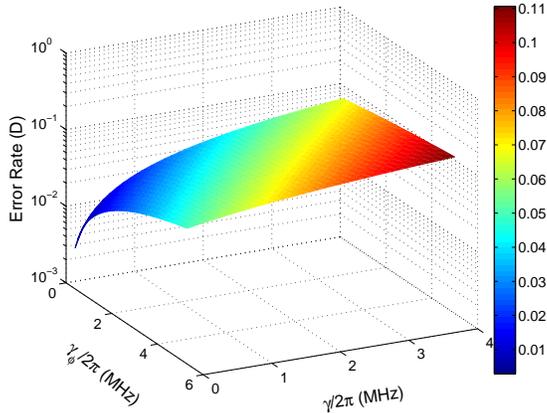,width=8cm} \caption{(Color Online) Dependence
of the error probability $D$\ in
entanglement generation on dissipation factor $\protect\gamma _{\protect%
\varphi }$ and $\protect\gamma $.}
\end{figure}

In conclusion, we propose a dispersive coupling scheme and find a
general interaction Hamiltonian for $n$ double-dot qubits and the
TLR. These double-dot qubits are biased at the large detuning region
and the TLR is always empty and virtually excited. The energy loss
of the cavity influences little on the quantum coherence of this
system. After analyzing the main decoherence sources, we show that
that various quantum operations of scalable solid-state quantum
computing could be reliably implemented with current technology.

This work was funded by National Fundamental Research Program, the
Innovation funds from Chinese Academy of Sciences, NCET-04-0587, and
National Natural Science Foundation of China (Grant No. 60121503, 10574126,
10604052).


\end{document}